\input{aipcheck}

\documentclass[
    ,final            
  ]
  {aipproc}

\layoutstyle{6x9}

\begin{document}

\title{Application of initial data sequences to the
study of 
Black Hole dynamical trapping horizons}


\newcommand*{\AEI}{Max-Planck-Institut f\"ur Gravitationsphysik,
  Albert-Einstein-Institut, Am M\"uhlenberg 1, D-14476 Golm, Germany}
\newcommand*{\MEU}{Laboratoire Univers et Th\'eories (LUTH), UMR 8102
  du C.N.R.S. Observatoire de Paris, Universit\'e Paris Diderot,
 F-92190 Meudon, France}
\newcommand*{\IAA}{Instituto de Astrof\'{\i}sica de Andaluc\'{\i}a, CSIC, Apartado Postal 3004, Granada 
        18080, Spain}

\author{Jos\'e Luis Jaramillo}{
  address={\IAA} , altaddress={\MEU}
}

\author{Marcus Ansorg}{
  address={\AEI}
}

\author{Nicolas Vasset}{
  address={\MEU}
}

\begin{abstract}
Non-continuous "jumps" of Apparent Horizons occur generically in
3+1 (binary) black hole evolutions.
The dynamical trapping horizon framework suggests a spacetime
picture in which these "Apparent Horizon jumps" are understood as
spatial cuts of a single spacetime hypersurface foliated by
(compact) marginally outer trapped surfaces. We present here
some work in progress which makes use of uni-parametric sequences
of (axisymmetric) binary black hole  initial data for
exploring the plausibility of this spacetime picture.
The modelling of Einstein evolutions by sequences
of initial data has proved to be a successful methodological tool
in other settings for the understanding of certain qualitative features
of evolutions in restricted physical regimes.
\end{abstract}

\maketitle

\paragraph{Problem and antecedents}
We aim here at gaining some insight into the understanding of non-continuous
Apparent Horizon {\em jumps} occurring generically 
in 3+1 evolutions of black hole spacetimes, with a focus on the binary case.
We look at this problem from the perspective of a geometric quasi-local
approach to black holes, namely the dynamical trapping horizons
introduced by Hayward and Ashtekar $\&$ Krishnan (cf. \cite{Haywa08}
for a recent review).
This framework suggests a spacetime picture in which these Apparent Horizon jumps 
actually correspond to different (non-connected) spatial cuts of a unique underlying
smooth hypersurface ${\cal H}$ foliated by (compact) marginally
outer trapped surfaces, when ${\cal H}$ is sliced by a given spacelike hypersurface 
$\Sigma$.
Our goal here is to assess this qualitative spacetime
picture by means of some specific numerical implementations.

\begin{figure}
  \includegraphics[height=.18\textheight]{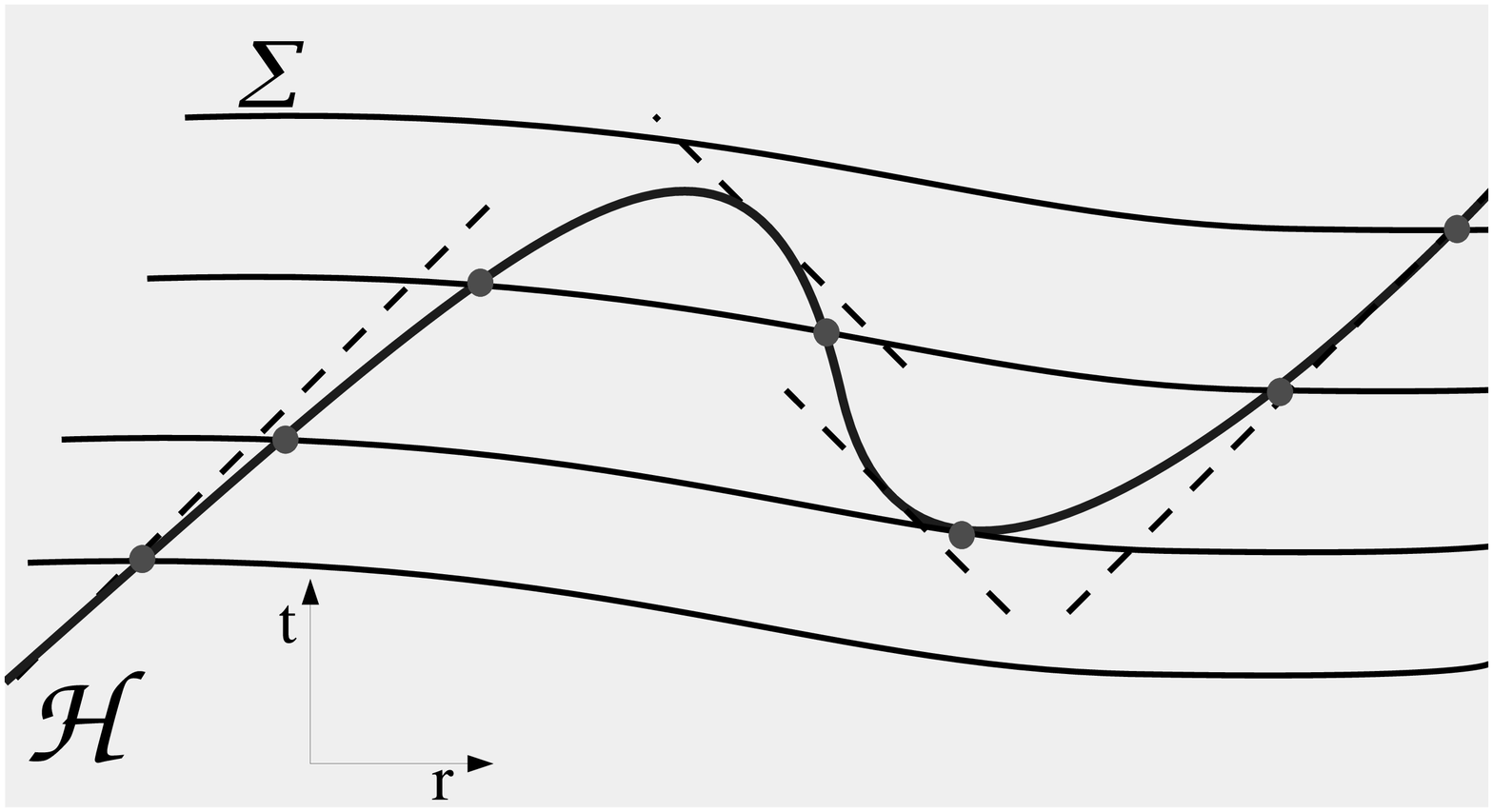}
\label{f:fig1}
  \caption{Apparent Horizons {\em jumps} in a  multiply sliced metric-type changing 
horizon ${\cal H}$.}
\end{figure}

\begin{figure}[h]
  \includegraphics[height=.15\textheight]{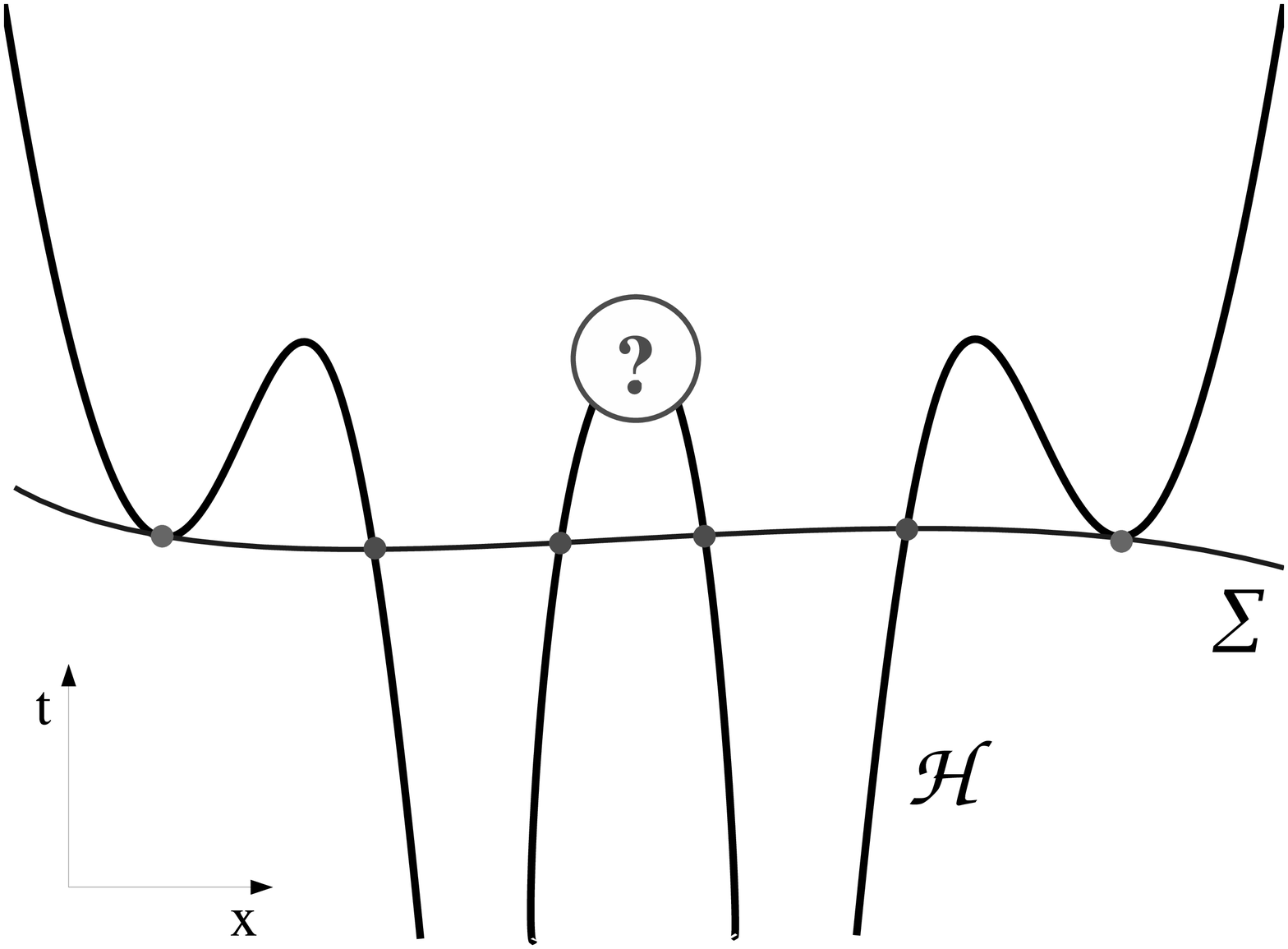} 
\hspace{0.5cm}
\includegraphics[height=.15\textheight]{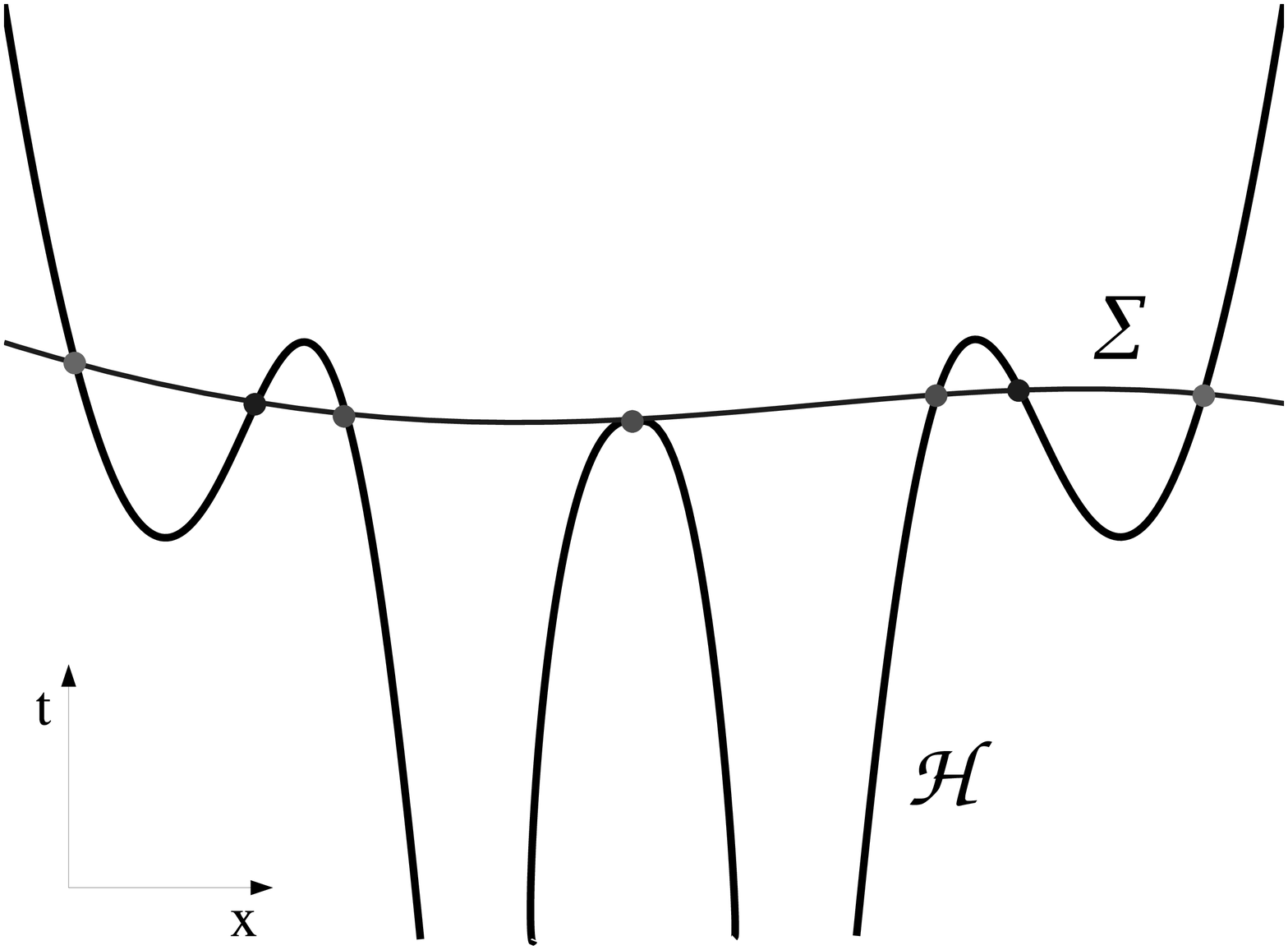}
\hspace{0.5cm}
\includegraphics[height=.15\textheight]{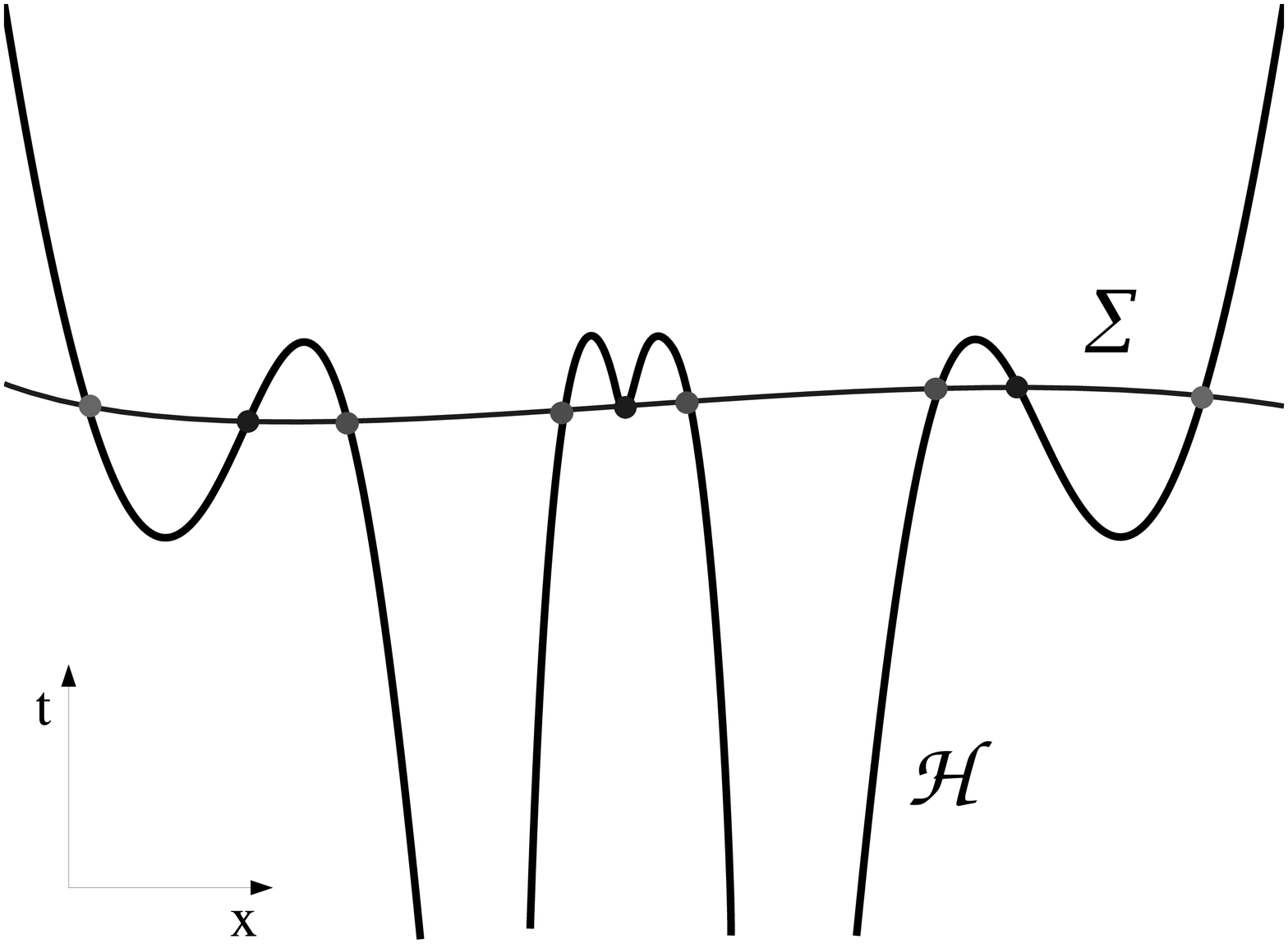}
\label{f:fig2}
  \caption{Some possibilities for the merger of the original and common inner horizons.}
\end{figure}

The setting of the dynamical trapping horizon framework is the quasi-local characterization
of a black hole in terms of a {\em trapped region}. 
The latter is built on the notion of {\em trapped surface},
a closed surface on which the expansions associated with {\em outgoing} $\ell^+$
and {\em ingoing} $\ell^-$ null congruences are negative: $\theta_+<0$ and $\theta_-<0$, respectively. 
The black hole horizon is then modelled as a worldtube ${\cal H}$ of marginally outer
trapped surfaces (MOTS), i.e. $\theta_+=0$. Additional geometric conditions are needed for 
${\cal H}$ to fulfill the expected physical properties of the horizon, namely:
a) a {\em future} condition, $\theta_-<0$, crucial for deriving an area growth law; and b)
an {\em outer} condition $\delta_{\ell^-} \theta_+<0$, meaning that when moving inwards the horizon 
we enter into the trapped region. Under the null energy condition, 
horizon ${\cal H}$ is either a spacelike or null hypersurface.
From a 3+1 perspective, we consider the intersections ${\cal S}_t$ of ${\cal H}$
with slices $\{\Sigma_t\}$ in a 3+1 spacetime foliation. A given $\Sigma_t$
can multiply intersect ${\cal H}$ giving rise to {\em Apparent Horizon} jumps
in the 3+1 description. In particular, this is the case if 
the dynamical horizon is actually part of a smooth MOTS-worldtube changing
its metric type (signature) along the evolution (i.e. if $\delta_{\ell^-} \theta_+>0$). 
This is illustrated in Fig. \ref{f:fig1} for the spherically symmetric case, where
$\pm 45^o$ lines correspond to null directions.
The suggested picture is that of a single MOTS-worldtube {\em bending} in spacetime. 
Focusing on the binary case, this translates into a scenario where a single smooth
MOTS-worldtube would account for every MOTS found in the $\Sigma_t$ slices.
If such a picture is correct then, starting from some initial slice with two
(non-connected) MOTS ${\cal S}_1$ and ${\cal S}_2$, once a common Apparent Horizon appears it
should immediately split into a growing common outer horizon
${\cal S}_{\mathrm co}$ (the Apparent Horizon) and a shrinking common
inner horizon ${\cal S}_{\mathrm ci}$. The outer one ${\cal S}_{\mathrm co}$ 
should asymptote to the Event Horizon whereas
the inner one ${\cal S}_{\mathrm ci}$ should somehow 
{\em annihilate} itself with the original horizons ${\cal S}_1$ and ${\cal S}_2$. Figure \ref{f:fig2}, 
where $x$ indicates the direction joining both black holes,
illustrates this process. Picture on the left represents the moment in which the common horizon
appears for the first time, whereas central and right pictures illustrate two possibilities
for the merging of ${\cal S}_{\mathrm ci}$ with  ${\cal S}_1$ and ${\cal S}_2$. In the central
one,   ${\cal S}_1$ and ${\cal S}_2$ first merge and the resulting MOTS gets {\em annihilated} with
 ${\cal S}_{\mathrm ci}$. In the picture on the right, ${\cal S}_{\mathrm ci}$ first pinches-off 
into two
MOTSs that annihilate respectively with  ${\cal S}_1$ and ${\cal S}_2$.
Of course, these or related ideas have already been discussed in the literature. 
We refer explicitly to recent references \cite{BooBriGon05,SchKriBey06,SziPolRez07}
where analytical and numerical studies of MOTS-worldtubes are carried out in different scenarios.
We also refer to the recent review \cite{Haywa08} where the issue of the signature change in ${\cal H}$
is discussed.
Results in \cite{SziPolRez07} are specially significant in the present context
since an overlapping of ${\cal S}_1$ and ${\cal S}_2$ is found in the dynamical evolution. This
explicitly challenges the single hypersurface picture suggested by dynamical horizons. 
Our main purpose is to ellucidate if a generic picture for the evolution of the inner horizon 
exists at all and, in that case, to assess if it corresponds rather to {\em pinchings} or 
{\em crossings} of the horizons.

\paragraph{Methodology}
The considered problem involves an intrinsicly evolutive process.
We lack a sufficiently accurate evolution code, so
we rather follow an alternative methodology by constructing
a sequence of data $(\gamma_{ij}, K^{ij})$. We stay in axisymmetry (MOTS are then axisymmetric)
and use spectral methods implemented in bispherical
coordinates. An excision technique is employed and a negative outer expansion $\theta_+= \theta_+^o <0$ 
is  prescribed
at the excision surface. Highly accurate snapshots are thus produced.

A criterium must be adopted to define the initial data sequence.
We assume that, once the common horizon has formed, the horizon
rapidly settles down to stationarity (radiation rapidly falls into the horizon or is radiated
away). Our sequences are thus only physically meaningful once the common horizon has already formed
and are characterised by the constancy of the ${\cal S}_{\mathrm co}$ area.
In practice, we impose the constancy of the ADM mass along the sequence and use
the ${\cal S}_{\mathrm co}$ area-constancy as a check.

\begin{figure}
  \includegraphics[height=.16\textheight]{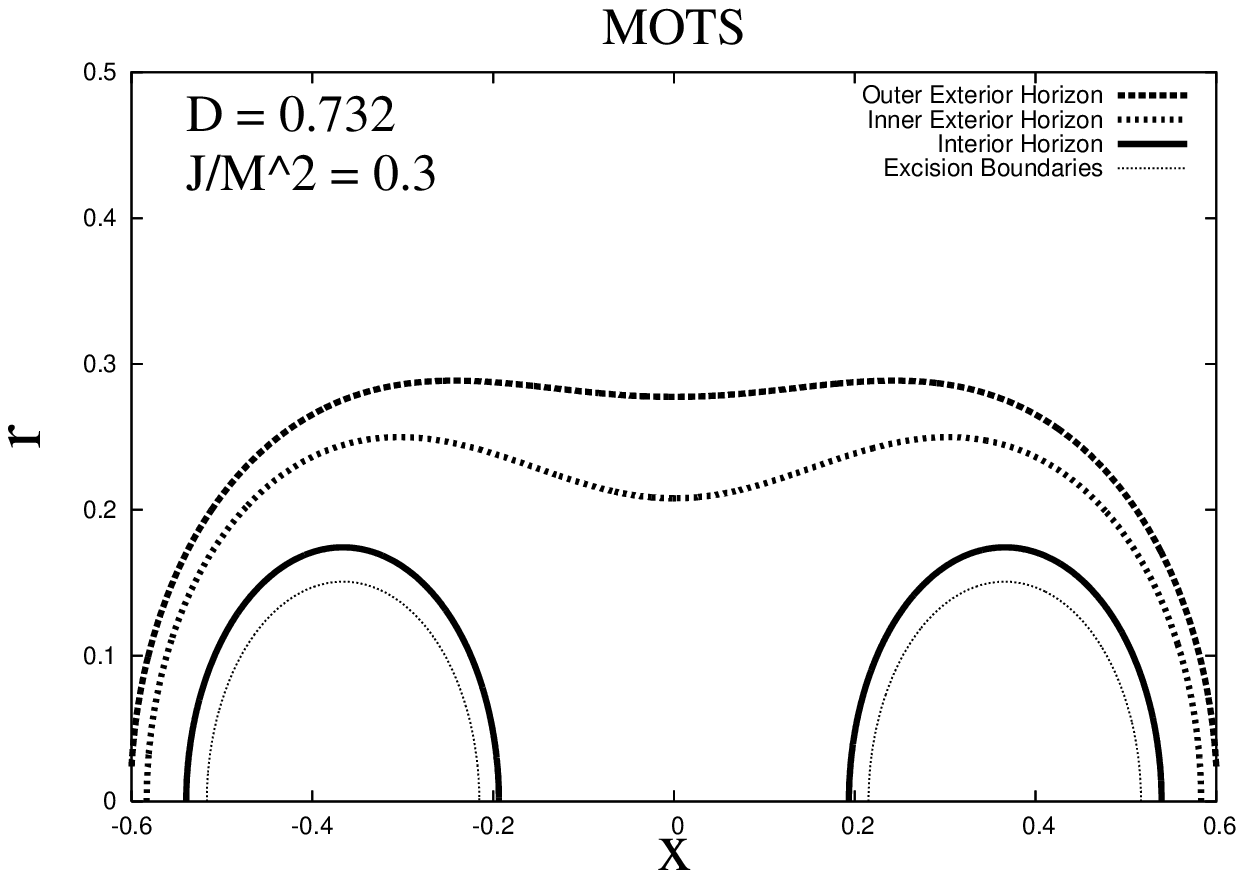} 
\hspace{0.1cm}
\includegraphics[height=.16\textheight]{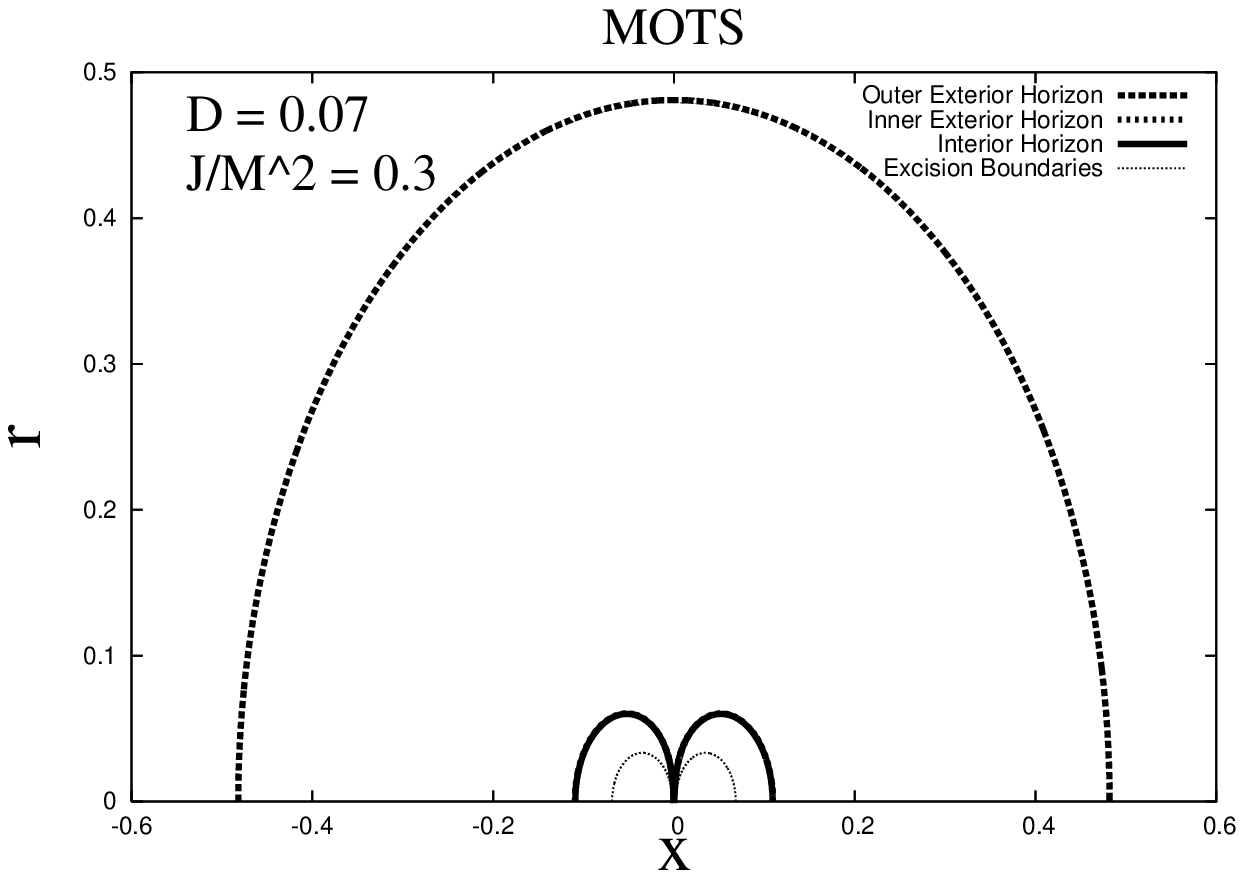}
\hspace{0.1cm}
\includegraphics[height=.16\textheight]{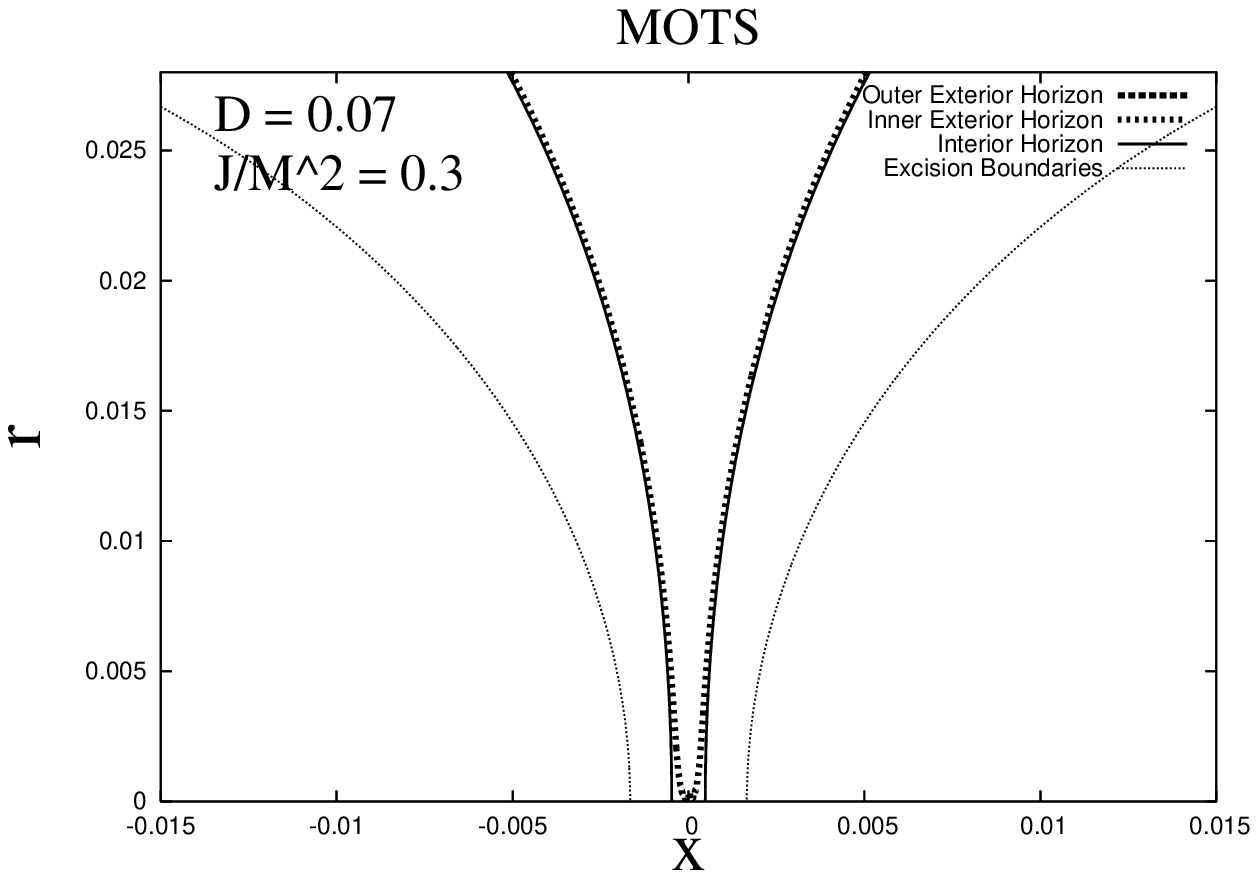}
\label{f:fig3}
  \caption{
Left: splitting of the common horizon in the $J/M^2=0.3$ non-boosted sequence.
Center:
accumulation of ${\cal S}_\mathrm{ci}$  and ${\cal S}_{1,2}$ at small $D$. Right:
central-part zoom of the $D=0.07$ sequence element.}
\end{figure}

\paragraph{Results}
We start by constructing non-boosted Bowen-York data, for two black holes rotating around the
axis joining them. We construct a uni-parametric sequence whose elements are labelled by 
the coordinate distance parameter $D$ between the coordinate centers of the excised surfaces.
The sequence starts at the value $D=D_o$ at which the common horizon appears for the first time. 
Then the sequence is constructed by disminishing $D$.
Proper distance between the horizons also disminishes, as checked {\em a posteriori}.
First, the common horizon splits (smoothly) into outer and inner MOTSs, 
consistently with the existence of a single MOTS-worldtube. As $D$ gets down, the common inner 
MOTS ${\cal S}_\mathrm{ci}$ approaches the original horizons  ${\cal S}_1$ and ${\cal S}_2$,
producing the kind of pictures found in \cite{SchKriBey06}.
However, when pushing down $D$ we observe none of the
expected possibilities (cf. Fig. \ref{f:fig3}). 
More specifically: i) original horizons
${\cal S}_1$ and ${\cal S}_2$ do not develop an {\em eight-shape} 
(i.e. they do not seem to merge) and ii) neither a pinching-off of the common inner horizon
${\cal S}_{\mathrm ci}$ is observed (previous experience
with Dyson-ring sequences, where a pinching does occur, strongly supports this statement). 
Moreover, no crossing/ovelapping of ${\cal S}_1$ and ${\cal S}_2$ like in \cite{SziPolRez07} 
is either observed. However, when trying to assess this last point, an important
{\em non-genericity} feature of our data becomes apparent: due to the vanishing
of  $K_{ij}s^i s^j$ and $K$, where $s^i$ is the normal to the MOTS, the considered MOTS are actually
minimal surfaces, $0= \theta_+ = D_i s^i + K_{ij}s^i s^j - K = D_i s^i$.
A maximum principle then applies: if two
such minimal surfaces tangentially touch, they must actually coincide.
No crossing can then happen. However, this {\em geometric obstruction} is just an artifact 
of the non-generic nature of the constructed data. 

In order to test the {\em crossing} possibility, we construct head-on {\em boosted} Bowen-York data,
for which the relevant components of the extrinsic curvature are non-vanishing.
The resulting sequence shows a non-trivial intertwining of MOTS and minimal 
surfaces. However when pushing $D$ down, the qualitative picture 
does not change with respect to the non-boosted case in Fig. \ref{f:fig3}: neither pinching 
nor crossing are observed, but rather the original horizons ${\cal S}_1$ and ${\cal S}_2$
accumulate against the common inner horizon ${\cal S}_{ci}$.

Geometric tests on the sign of $\theta_-$ and $\delta_{\ell^-}\theta_+$ 
(more precisely on  $-\delta_{s}\theta_+$) show that the common outer 
horizon behaves as a dynamical horizon, whereas the common inner one
violates the {\em future} and {\em outer} geometric conditions.

\paragraph{Conclusions and perspectives} Results in \cite{SziPolRez07}
challenge the single smooth MOTS-worldtube picture for Black Hole
coalescence suggested by the dynamical trapping horizon framework.
We have not found numerical evidence for:
a) neither interior horizons annihilation (single smooth MOTS-horizon picture)
nor b) {\em crossing} of the original horizons \cite{SziPolRez07}.
As a third possibility, there could exist three distinct never-merging MOTS-worldtubes, namely
an exterior one accounting for the two common horizons, and two other
worldtubes corresponding to the original black holes.
In sum, the existence of a generic picture is seriously challenged.
The whole problem could be reformulated and {\em refined} as
the search of conditions on the 3+1 slicing guaranteeing the existence of a unique underlying 
smooth MOTS-worldtube ${\cal H}$.
The main motivation for this is to set {\em geometric flow equations} along ${\cal H}$
relating quantities before and after the coalescence. Such equations
explicitly need the existence of some underlying smooth hypersurface.

There are some important {\em caveats} in our discussion. Namely, we are not
dealing with an actual evolution, but only with a sequence of
snapshots. One would need to assess the error in the evolution Einstein equations.
On the other hand, the fact that we have not been able to construct a single initial data 
showing either {\em pinching} or {\em crossing} deserves attention. Of course, genericity issues
regarding our data must still be addressed.

Regarding future perspectives, the main interest of this research line
is geometric, aiming at understanding the dynamics of MOTS-worldtubes. 
In this sense, numerical relativity simulations  
can be crucial for establishing  {\em generic}
behaviours. This can provide relevant insights 
in the gravitational collapse context, e.g. Penrose inequality
or more generally cosmic censorship conjecture, as well as in the study of
the Hawking radiation process when taking into account corrections
to General Relativity.

\bibliographystyle{aipproc}

\end{document}